\RequirePackage{fix-cm}
\documentclass[smallextended]{svjour3}       
\smartqed  
\usepackage{booktabs}
\usepackage{graphicx}
\usepackage{xcolor}
\usepackage{fancybox}
\usepackage{pdflscape}
\usepackage{caption}
\usepackage{subcaption}
\usepackage{tabularx}
\usepackage{longtable}
\usepackage{array}
\usepackage{fontawesome}
\usepackage{multirow}
\usepackage{array}
\usepackage{easyReview}
\usepackage[justification=centering]{caption}

\def\mybarhhigh#1#2{
   {\color{black}\rule{#1mm}{4pt}}  #2}

\usepackage{wrapfig}
\usepackage{subcaption}
\usepackage{booktabs}
\usepackage{setspace}
\usepackage{tikz}
\usepackage{amssymb}
\usepackage{booktabs}
\usepackage{enumitem}
\usepackage{changepage}
\usepackage{hyperref}
\usepackage[T1]{fontenc}
\hypersetup{
    colorlinks=true,
    linkcolor=black,
    filecolor=black,      
    urlcolor=black,
    citecolor=black,
}
\usepackage{graphicx}

\usepackage{fancybox}
\usepackage{longtable}
\usepackage{tasks}
\usepackage{pdfpages}

\begin{document}

\title{Engineering Adaptive Information Graphics for Disabled Communities: A Case Study with Public Space Indoor Maps}

\titlerunning{Engineering Adaptive Graphics for Disabled...}        

\author{Anuradha Madugalla        \and
        Yutan Huang \and John Grundy \and Min Hee Cho \and Lasith Koswatta Gamage \and   \and Tristan Leao \and Sam Thiele 
}


\institute{A. Madugalla\at
              Dept. of Software Systems and Cybersecurity, Monash University, Melbourne, Australia \\
              \email{anu.madugalla@monash.edu} 
           \and
        Y. Huang\at
              Dept. of Software Systems and Cybersecurity, Monash University, Melbourne, Australia \\
              \email{yutan.huang@monash.edu}  
           \and
           J. Grundy \at
              Dept. of Software Systems and Cybersecurity, Monash University, Melbourne, Australia \\
              \email{john.grundy@monash.edu}
              \and
            M.H. Cho \at
              Dept. of Software Systems and Cybersecurity, Monash University, Melbourne, Australia \\
              \email{minhee.c.j@gmail.com}       
           \and
             L.K. Gamage\at
              Dept. of Software Systems and Cybersecurity, Monash University, Melbourne, Australia \\
              \email{lasithvindu1@gmail.com} 
           \and
            Y.P. Lau\at
              Dept. of Software Systems and Cybersecurity, Monash University, Melbourne, Australia \\
              \email{desmondyplau@gmail.com}         
           \and           
           T. Leao \at
              Dept. of Software Systems and Cybersecurity, Monash University, Melbourne, Australia \\
              \email{tlea0003@student.monash.edu}
            \and           
           S. Thiele \at
              Dept. of Software Systems and Cybersecurity, Monash University, Melbourne, Australia \\
              \email{sthi0002@student.monash.edu}
}

\date{Received: date / Accepted: date}

\maketitle

\begin{abstract}
Most software applications contain graphics such as charts, diagrams and maps. Currently, these graphics are designed with a ``one size fits all" approach and do not cater to the needs of people with disabilities. Therefore, when using software with graphics, a colour-impaired user may struggle to interpret graphics with certain colours, and a person with dyslexia may struggle to read the text labels in the graphic. Our research addresses this issue by developing a framework that generates adaptive and accessible information graphics for multiple disabilities. Uniquely, the approach also serves people with multiple simultaneous disabilities. To achieve these, we used a case study of public space floorplans presented via a web tool and worked with four disability groups: people with low vision, colour blindness, dyslexia and mobility impairment. Our research involved gathering requirements from 3 accessibility experts and 80 participants with disabilities, developing a system to generate adaptive graphics that address the identified requirements, and conducting an evaluation with 7 participants with disabilities. The evaluation showed that users found our solution easy to use and suitable for most of their requirements. The study also provides recommendations for front-end developers on engineering accessible graphics for their software and discusses the implications of our work on society from the perspective of public space owners and end users. 

\end{abstract}

\section{Introduction}
Information graphics such as charts, diagrams, and maps are used in software applications to convey complex information concisely and in an easy-to-understand format \cite{burmark2002visual}. For this reason, they are heavily used in software focusing on fields such as entertainment, education, health, and finance. However, these are currently designed for an "average" user and do not address the specific needs of minority groups, such as disabled users who represent 15\% of the world population \cite{Disabili33_online}. Therefore, these groups face difficulties in accessing the information graphics. Disabled communities can be people with sensory and speech impairments (blind, deaf), intellectual impairments (cognitive impaired, dyslexia), physical impairments (wheelchair users, motor impaired) and more. Researchers have identified this problem and have developed solutions to generate more accessible graphics for different disabled groups. These include graphics presented with audio cues and 3D graphics for the vision impaired \cite{butler2021technology,madugalla2020creating}, multi-modal graphic presentation with audio and gestural information for dyslexia \cite{pollak2005dyslexia}, and using shapes, and patterns instead of colours on graphics for the colour impaired \cite{R13}. However, all these solutions focus only on a single disability, and to the best of the authors' knowledge, there are no solutions that cater well enough for multiple disabilities. 

In this research, we aim to address multiple disabilities using the concept of adaptivity. Adaptivity has traditionally focused on device-driven characteristics like size, format, and screen resolution \cite{Adaptive2_online}. Recently, UI research has also started to adapt UI layout, themes, and textual content \cite{gustafson2015adaptive,luy2021toolkit}. However, adaptivity for disabilities is yet to be explored for graphics. The lack of focus on graphic adaptivity can also be attributed to the challenges in adapting the currently most used graphic format, raster. We overcome this barrier by focusing on the other most common graphic type, vector. We specifically concentrate on Scalable Vector Graphics (SVG), as this is a standard graphics format. 
This research presents an approach to designing accessible graphics for multiple disabilities using adaptive SVG graphics. It involves requirements gathering, development and a preliminary evaluation with disabled users to achieve this. In this study, we specifically focus on four types of disabilities to achieve this. These are,
\begin{enumerate}
    \item \textit{Low Vision}: People who have permanent vision loss that cannot be corrected with prescription glasses
    \item \textit{Colour Impairment}: People who struggle to perceive colour differences due to complete or partial loss of cone systems
    \item \textit{Dyslexia}: People who struggle with tasks associated with language processing: reading, spelling and associating speech patterns to letters
    \item \textit{Mobility Impairment}: People who experience a permanent or temporary reduction in their ability to navigate their environment
\end{enumerate}
We chose these categories as they are some of the most common disabilities \cite{Disabili8_online} and as authors had pre-established connections with organisations that work with these individuals.  

The rest of the paper is structured as follows. In the next section, we present a motivating example of our work, followed by related work. The section 4 discusses our approach for generating accessible graphics and sections 5, 6 and 7 contain the three steps of this approach: requirement elicitation, system design, and development/implementation, respectively. Section 8 includes a preliminary evaluation, Sections 9 and 10 have threats to validity and recommendations, respectively, and the last section summarises the paper while presenting suggestions for future work.

\section{Motivation}
\begin{figure}[h]
\includegraphics[width=\textwidth ]{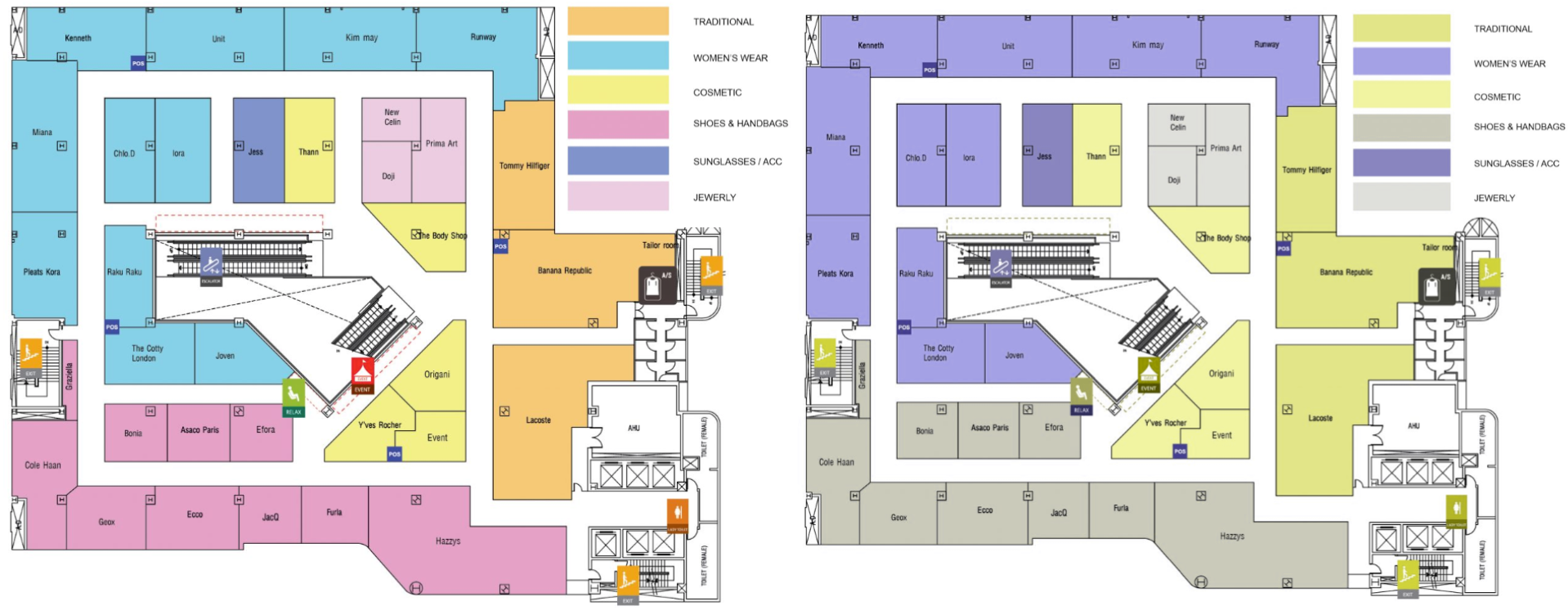}
\caption{A floorplan diagram (a) Original graphic (b) View of a person with Deuteranopia (Red-Green colour blindness)}
\label{fp_colorblind}
\end{figure}

Consider the example of an indoor map from a retail centre website as shown in Figure \ref{fp_colorblind} (a). Currently, this map is designed for an average user by following a ``one size fits all" approach and ignores the specific characteristics and capacities of many people with disabilities. For instance, 1) someone with red-green colour blindness won't be able to separate most colour-coded shop categories in this map (Figure \ref{fp_colorblind} (b), 2) a low vision user will struggle to read the small fonts used for the store names, 3) a person with dyslexia would face difficulties in reading the used font family, 4) for a wheelchair user, the issues will be around the lack of critical information such as the location of lifts, ramps and corridor widths. These problems limit their access to information, making it challenging to navigate these spaces independently. This is discrimination against disability and a violation of laws in many countries \cite{Disabili8_online}. Due to this importance of public space maps, we decided to use the case study of indoor maps in our study. 

\subsection{Key Challenges}
Researchers have sought to address this challenge by developing accessible graphics based on each disability. This has led to a collection of standalone solutions with their own device and platform constraints, making it challenging to combine these. Therefore, there is a need for a comprehensive solution that addresses multiple disabilities, ideally in a single graphic. One challenge in this approach is that it may lead to an overly complicated graphic that is cluttered and complex. A second challenge is a need for compromises when requirements between groups clash with each other, e.g. colour preferences change between different types of colour-blinded users; people with ADHD may prefer layered information with less information in each layer, but someone with memory issues such as elderly may choose less layered information. Therefore, while it is imperative to address the various needs of disabled users, this can not be achieved with a single graphic in a conventional sense. Our research seeks to address this problem by developing an approach that supports integrating complex requirements into the SVG's XML back-end and reveals only the necessary information per group adaptively. In doing this, we plan to answer the following research questions:

\begin{enumerate}
    \item \textbf{RQ1}: What are the requirements from an inclusive information graphic?
    \item \textbf{RQ2}: How to address these requirements adaptively via a single solution? 
    \item \textbf{RQ3}: To what extent does the disabled community accept our adaptive graphics solution?
\end{enumerate}

\section{Background}

\subsection{Graphic Accessibility} 
Many researchers have explored accessible graphics for users with vision issues. They have devised solutions that involve different combinations of audio cues, tactile displays, 3D models and vibration to make graphics accessible for legally blind users \cite{butler2021technology,madugalla2020creating,calle2020designing}. These technologies help users explore graphics with their hands, allowing them to build a mental graphic model. Colour-impaired users are also addressed under graphic accessibility, and these solutions focus on presenting a selection of colour scheme options for colour-blind users to choose from \cite{R12}, and using information other than colour to present visual information, such as shapes, sizes, and patterns \cite{R13}. 

Accessible graphics are explored for people with dyslexia, where studies explore using multi-modal graphics presentations, such as audio, gestural information, and written language, on top of traditional visual representations \cite{pollak2005dyslexia}. Another group of users that benefit from accessible graphics are people with mobility impairment, which involves motor impairment, wheelchair users, cane users, crutch users and others whose reduced mobility presents difficulties in performing everyday physical tasks such as interacting with graphics \cite{sarsenbayeva2022methodological}. Independent tools like head-based point trackers and eye-based interaction graphical user interfaces (GUI) have been developed to help motor-impaired users. However, these tools are yet to be explicitly implemented for graphics \cite{istance1996providing,cicek2020designing}. Most of these involved developing prototypes and evaluating with end-users via surveys, interviews and focus groups. While there are all these individual solutions, there is still no solution that caters to multiple disabilities.


\subsection{Map Accessibility} 
Maps are essential for navigating unfamiliar spaces \cite{engel2020travelling}. Map-based navigation solutions have focused on various disabilities, such as vision-impaired, cognitive-impaired, and mobility-impaired \cite{zahabi2023design,rink2023walking}. 
In these studies, based on the focused user group, the supported interaction methods changed significantly. For wheelchair users, solutions supported enabling voice commands and avoiding manual interaction to free the hands to push their wheelchairs \cite{barbosa2018trailcare}. Audio and tactile cues were used in interactions with blind users \cite{madugalla2020creating}. For the hearing impaired, a combination of visual and tactile feedback was provided \cite{rodriguez2017gawa}. 

The outdoor maps in these studies used GIS data, such as OpenStreetMap and PostGIS, to directly build accessible maps \cite{taylor2016customizable,R17}. Indoor maps use Bluetooth beacons and WiFi to extract user locations in buildings and map them to floorplans provided by buildings \cite{chang2008mobile}. However, the accuracy of indoor map localisation requires significant improvement \cite{zahabi2023design}. Due to these reasons, there is less work on indoor maps than on outdoor maps. Most of this research involved a design guideline generation stage based on existing application reviews or via direct interactions with end-users. Many recruited end-users of the target group and conducted usability evaluations to validate their concepts or applications. The predominant method employed for these assessments was user or field testing, during which end-users were tasked with using the interface to navigate indoor or outdoor settings \cite{zahabi2023design}. The evaluator then assessed their performance and gathered their subjective feedback on the solution, typically through surveys or interviews conducted at the end of the study. This work needs to be extended to cater to multiple disabilities. 

\subsection{Adaptivity }

Adaptivity is used by many researchers when different types of requirements need to be addressed in a single solution. It is used in areas such as User Interface (UI) design to adapt content based on human needs and features. There have been studies on developing adaptive UI for chronic diseases \cite{awada2018adaptive,wangadaptive}, age \cite{machado2018conceptual,gregor2002designing}, vision impaired \cite{iqbal2018usability,khan2018blindsense} and on multiple vulnerable groups, such as colour impairment, low vision and dyslexia \cite{luy2021toolkit}. However, all these different forms of adaptivity focus on web elements such as textual content, layout, and colour in UI, and none has focused on making graphics adaptive.  

This can be attributed mainly to the challenges in accessing and editing the currently most used graphic format: raster. Raster (PNG, JPG) uses pixel-based representation, meaning pixel-level changes must be made to implement adaptivity on any graphical elements, such as layout, colour, and fonts \cite{madugalla2020creating}. This requires advanced image processing mechanisms, and even after using such methods, the graphic can still be distorted due to the granular, pixel-level changes \cite{marriott2002fast,Rastervs52_online}. A possible solution is to explore adaptivity with vector graphics. 

\subsection{Scalable Vector Graphics (SVG)} 
SVG is a type of vector graphic that uses a set of XML-based tags to position lines, shapes and other geometric features on a plane. Compared to raster, vector such as SVG has several advantages, such as faster download speeds due to reduced size, and the ability to scale in high resolution without allowing distortions \cite{marriott2002fast,eisenberg2014svg}. In generating adaptive graphics, SVG's XML-based back-end will support direct manipulation of the graphic via its tags and allow modification of colors, and fonts via tag attributes. 

Researchers focusing on vision-impaired users have started exploring SVGs to generate accessible graphics. In one solution, using SVG's XML back-end, a colour-filled graphic was converted to a simple line graphic, which was then directly printed as a tactile graphic \cite{krufka2007visual}. Another embedded supplemental information within the SVG tags of a map, which was then interpreted by assistive technology devices such as screen readers \cite{R14}. However, SVG is yet to be explored to generate adaptive graphics for multiple disabilities.


\newpage

\section{Our Approach}

\begin{figure}[h]
\includegraphics[width=\linewidth, trim={0.5cm 0.5cm 0.5cm 0.5cm},clip]{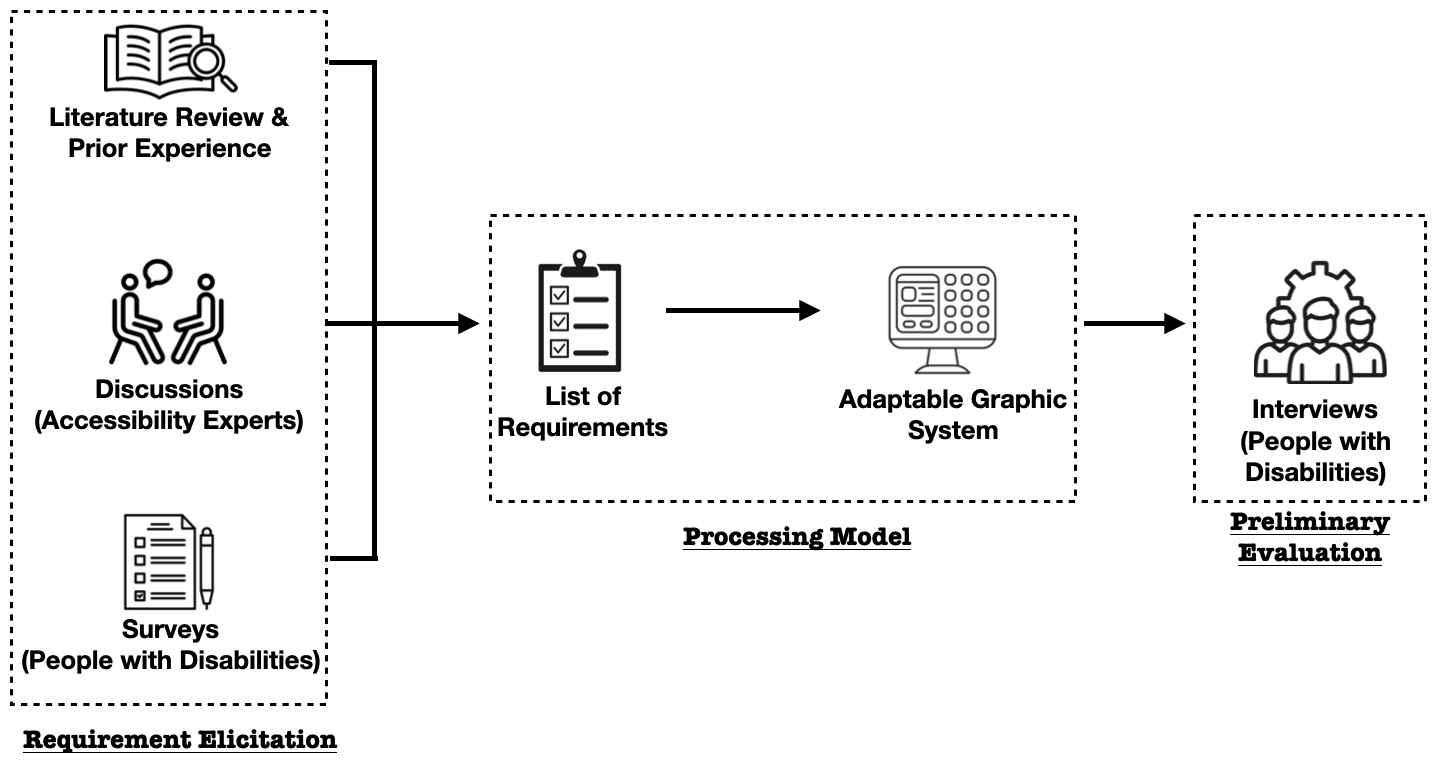}
\caption{Research design} 
\label{fig_researchDesign}
\end{figure}

Our study was conducted in three stages, as shown in Figure \ref{fig_researchDesign}: requirements elicitation, system development and a preliminary evaluation. Our requirements elicitation stage consisted of gathering information from a detailed literature review, interviewing three experienced accessibility experts, and conducting a survey with 80 disabled users. By obtaining requirements from both experts and disabled users, we sought to conduct a more rigorous requirement elicitation \cite{baez2018agile}. We used these requirements to create a novel SVG-based adaptable graphics generation system. The final prototype adaptive SVG-based map system was evaluated with a user study involving interviews with seven disabled users.  

\subsection{Case Study of Public Space Floorplans}
Based on this research design, we developed adaptive graphics for a public space floorplans case study. Public spaces encompass a large variety, including educational: libraries,  public transport: train stations, entertainment: museums, and commercial: resorts. Floor plans are used to pre-plan visits to these spaces and navigate once inside the building. As these are designed for average users, most people with disabilities face challenges in accessing them, which leads to issues in navigating these spaces independently. We applied our research design on these floorplans to generate accessible versions of these graphics.

\section{Requirements Elicitation}
Our requirement elicitation consisted of three steps: 1) reviewing related work, 2) interviewing representatives of key disability support groups, and 3) surveying disabled users. 

\subsection{Reviewing Related Work}
Not much work has been conducted on designing graphics for the multiple disabled groups. By analysing prior work from building environments and web design, we identified the following requirements/guidelines:
\begin{enumerate}[leftmargin=*]
    \item Highlight features that are deemed important for accessible design, such as ramps, tactile flooring, accessible toilets, and support desks \cite{RN1}
    \item Support multiple disabilities: Some may have more than one disability \cite{luy2021toolkit}
    \item Use accessible colour schemes \cite{R21}
    \item Use icons when possible instead of text \cite{muftah2020investigating}
    \item Keep unnecessary elements to a minimum \cite{R16}
    \item Group elements with a similar focus together \cite{R16}
    \item Keep everything on one page and reduce/avoid scrolling \cite{R16}
    \item Ensure map controls do not overlap the map: otherwise, it is challenging for visually impaired users to locate these \cite{R16}
\end{enumerate}

\subsection{Mock-ups and Expert Feedback}

\begin{figure}[h]
\centerline{\includegraphics[width=\columnwidth, trim={1cm 2.5cm 1cm 2.5cm},clip]{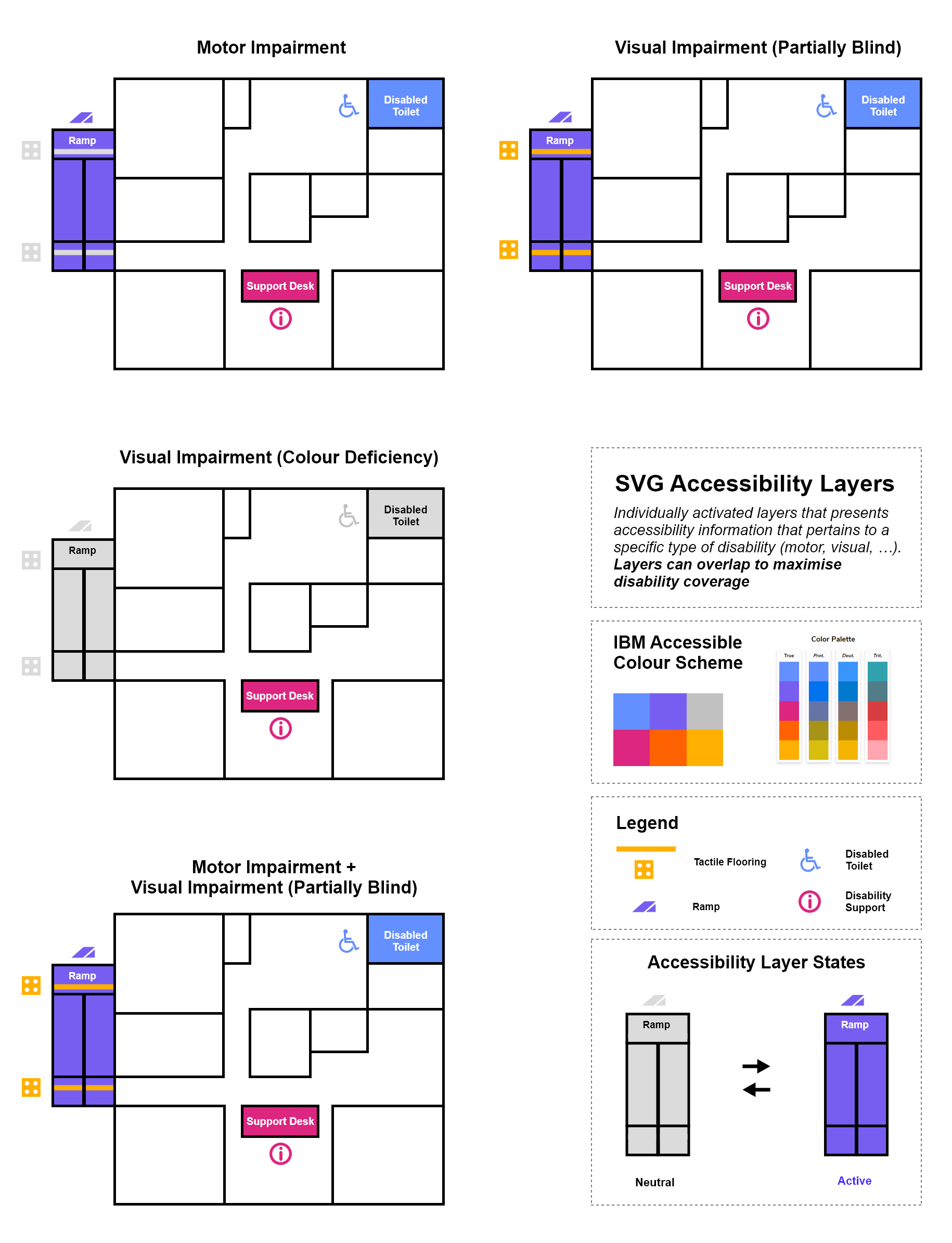}}
\caption{Mock-ups of Adaptive Floorplans}
\label{fig_mockup}
\end{figure}

Based on the guidelines from related work, we developed mock-ups, some of which are shown in Figure \ref{fig_mockup}. We presented users with individually activated layers where each layer had highlights for accessibility information related to one specific type of disability. Our mock-up supported multiple layers to overlap in order to support users with multiple conditions. We utilised the IBM accessible colour scheme as the default palette and used graphical icons as an alternative method of conveying information. 

To obtain feedback on our initial mock-ups, we conducted informal interviews with three executives from disability support groups. These were the manager of Monash University disability support services, an executive officer from AUSPELD, and an executive of Ability First Australia. The interviews involved presenting the mock-ups to the participants and asking for general feedback on our colour schemes, floorplan elements, and text labels. 

Their feedback indicated that these experts approved of our layer-based design and highlights based on disability. They did not perceive an issue in the colour schemes or icons used. However, they had suggestions on how these can be further improved, and these are listed as additional requirements below. 
\begin{enumerate}
    \item Adding more information about accessibility impediments such as heavy doors and staircases as these are essential for mobility and sight-challenged users
    \item Using an inclusive language: ``accessible" rather than ``disabled", when describing toilets
    \item Use simplified language in all user interactions
\end{enumerate}

Using the understanding gained from related work and these discussions, we designed and conducted a detailed survey of disabled users, as presented in the next section.

\subsection{User Survey}
\subsubsection{Survey Design}
We wanted to understand how people with disabilities used floorplans, and by asking questions about their issues and suggestions for improvements, we extracted their requirements for a floorplan graphic that adapts to their needs. A link to the detailed survey can be found here [\url{https://tinyurl.com/4kf6j4hv}]
\begin{enumerate}
    \item Demographics: Age, gender, country, disabilities, how long were they disabled, methods used for navigation support
    \item Experiences in public spaces: Types of public spaces they visited and how they planned these visits
    \item Floorplans: How frequently these were used, methods of accessing, issues in using and suggestions for improvements
\end{enumerate}

\subsubsection{Recruitment}
We recruited participants using the probability cluster sampling technique \cite{denscombe2017ebook}. We advertised our study on three main channels. One was via social media platforms, including social media groups of people with disabilities. The second channel was via organisations, namely AUSPELD (the Australian Federation of Specific Educational Learning Difficulties Associations), University disability support units, an eye hospital in China and a disability school in China. The third channel was advertising on Prolific, an online platform that helps researchers recruit participants from a pool of pre-registered participants. These multiple venues helped us obtain 80 responses, which is considered a significantly large data set when working with disabled users. 

\subsubsection{Data Collection}
According to Denscombe's categorisation, we chose the method of an internet survey which was launched as a web questionnaire \cite{denscombe2017ebook}. We used the Qualtrics platform for this. For the participants from the eye hospital and the disability school in China, we translated the survey into Mandarin. A representative at these locations obtained responses from participants in a face-to-face discussion and shared these responses with us. A research team member translated the answers into English and entered them into the Qualtrics platform. 

\subsubsection{Data Analysis}
We analysed our data using a mixed-method approach. We had 6 close-ended questions, mainly focusing on demographics and 11 open-ended questions focusing on experiences with public spaces and floor plans. We used descriptive statistics from the quantitative field for close-ended questions. For the open-ended questions, we used thematic analysis. We first used open coding to assign codes for each response, then categorised these codes and lastly, grouped them by themes. The results of these analyses are described below. 

\textbf{Demographics}
The demographic distribution of our participants is shown in Table \ref{tab_demographics}. Under disabilities, we have 1) Single disability: participants with one of our four chosen disabilities, and 2) Multiple disabilities: participants with more than one of these disabilities and possibly other disabilities. As shown in Table \ref{tab_demographics}, there were 34/80 participants with multiple disabilities. If a person had colour impairment along with another disability/ies, we did not count them in the colour vision category; instead, we included them in this category.

\begin{table}[h]
\centering
\caption{Demographics of Participants}
\begin{tabularx}{\columnwidth}{l r | l r | l r } 
 \toprule
 \textbf{Gender} &   & \textbf{Disability} &  & \textbf{Countries} &  \\ 
 \midrule
 Male & 37 & Low vision issues & 16 & China & 29 \\
 Female & 42 & Dyslexia & 12 & USA  & 12 \\
 Others & 1 & Mobility Impaired & 14 & UK & 12 \\
 \cmidrule(r){1-2}
 \textbf{Age} &  & Color Impaired & 4  & Sth Korea	& 9\\
\cmidrule(r){1-2}
\textless 30 & 44  & Multi. Disabilities & 34  & Australia & 8  \\
\cmidrule(r){3-4}
31-50 & 27 & \textbf{Mobility} & \textbf{Aids}   & Poland	& 3  \\
\cmidrule(r){3-4}
51+ & 9 & Canes & 12  &  Sth Africa & 2 \\
\cmidrule(r){1-2}
 \textbf{Duration} &  & Scooters & 7   &  Others & 5  \\
 \cmidrule(r){1-2}
\textless 2 Yrs. & 5  &  Wheelchairs & 5   &  &\\
 3-10 Yrs.& 22 &  Guides & 4   &  & \\
 11-20 Yrs.& 20 & Combined & 14   &  & \\
 21+ Yrs.& 32 &  &   &  &    \\
 \bottomrule
\end{tabularx}
\label{tab_demographics}
\end{table}

\textbf{Public spaces: } We found that our participants visited a range of public spaces, including shopping centres, supermarkets, parks, hospitals, etc. Out of these, shopping centres and supermarkets were the most visited. The distribution of visits between these places for 80 participants is shown in Table \ref{fig_visits}.  
\begin{table}[h]
\centering
\caption{Participants' Visits to Public Spaces}
\label{fig_visits}
\begin{tabular}{@{}llll@{}}
\toprule
\textbf{Type of Space} &   Often & Periodically  & Rarely \\ 
\midrule
Shopping Centres     & \mybarhhigh{5}{10} & \mybarhhigh{13.5}{27}  & \mybarhhigh{1}{2}\\
Supermarkets   & \mybarhhigh{12}{24} & \mybarhhigh{1.5}{3}  & \mybarhhigh{1}{2}\\
Parks   & \mybarhhigh{3.5}{7} & \mybarhhigh{2}{4}  & \mybarhhigh{0.5}{1}\\
Hospitals   & \mybarhhigh{2.5}{5} & \mybarhhigh{1.5}{3}  & \mybarhhigh{1}{2}\\
Libraries   & \mybarhhigh{1.5}{3} & \mybarhhigh{2}{4}  & \mybarhhigh{0.5}{1}\\
Museums   & \mybarhhigh{0.5}{1} & \mybarhhigh{1.5}{3}  & \mybarhhigh{2}{4}\\
\end{tabular}
\end{table}

34\% of our participants pre-planned their visits to these spaces, while 65\% did not, as they only visited familiar public spaces or had an accompanying friend/relative when visiting new places. Those who pre-planned searched for floorplans and other information using 1) online technology [websites, mobile apps, search engines and public reviews] and 2) human assistance [friends and relatives]. They used pre-planning to find ``shortest routes" and ``wheelchair access, lifts, exits and disabled access toilets". They considered pre-planning helpful to ``avoid and anticipate any challenges" and ``because visits [to unknown places] make me anxious". 

\textbf{Usage of maps:} We asked our participants if they used floorplans or maps during their visits to these complex public spaces. There was a close split between Yes and No responses, with a few stating they used floorplans very often. The reasons for using these were identified as a travel aid to navigate new places and to locate emergency exits. All our participants with colour blindness, low vision, or mobility impaired (wheelchair users) said they used floorplans often. For dyslexia, it was a split between yes and no. 

\textbf{Methods of accessing maps: } Most participants found floorplans onsite at building entrances, elevators, handouts and information desks. Some found them online from websites, via apps created for these spaces, and one mentioned emailing the organisations. Some used a combination of onsite and online methods, and a few did not use floorplans as they were hard to find. 


\textbf{Required information: }Most of our participants required information about the layout of the public spaces from the floorplans where they searched for locations of shops and toilets. Additionally, they searched for facilities such as elevators, ramps, entrances and exits, accessible parking areas, accessible seating, information desks, water filling stations, location and count of stairs, and terrain information. Many also searched for navigation information such as shortest paths, accessible paths or a general path to a destination. 

\textbf{Issues in current maps: } Most of our participants faced several problems in using the floorplans while some did not have any issues. Some identified 1) Technological issues such as bugs, not being updated, missing functionalities and general unavailability. But most reported 2) Design issues such as \textit{Font issues:} small fonts, hard to follow font styles; \textit{Colour issues: } non-colour blind friendly colour usages, confusing use of colours and personal dislike to adopted colour schemes; \textit{Information overload: }- overwhelming design, ``if there are massive legends ... or if there are a lot of semi-randomly coloured lines I kinda just skip over them in my head". Few also reported they found maps hard to understand as graphics are blurred for them, didn’t understand graphics in general, and 1 participant mentioned, ``I’m too old to understand the map". 

\textbf{Suggestions for improvements: }Key suggestions included three categories. Category 1 was design changes: \textit{colour changes} to use more accessible colours, clearly distinctive colours; \textit{font changes} to use larger fonts and different font styles, and \textit{enlarging} maps to fit user needs. Category 2 was Content changes: such as changing the level of information to simplify content to make it easier to read, replacing text with icons and using more straightforward vocabulary as well as adding more accessibility information and highlighting them, e.g. ramps, accessible toilets, emergency exits, size of doors etc. Category 3 was Technological changes: such as having regular updates and bug fixes, adding new features such as voice-over, intelligent voice assistants and using 3D maps.  

\subsection{Identified Requirements}
\begin{figure}[h]
\includegraphics[width=\linewidth]{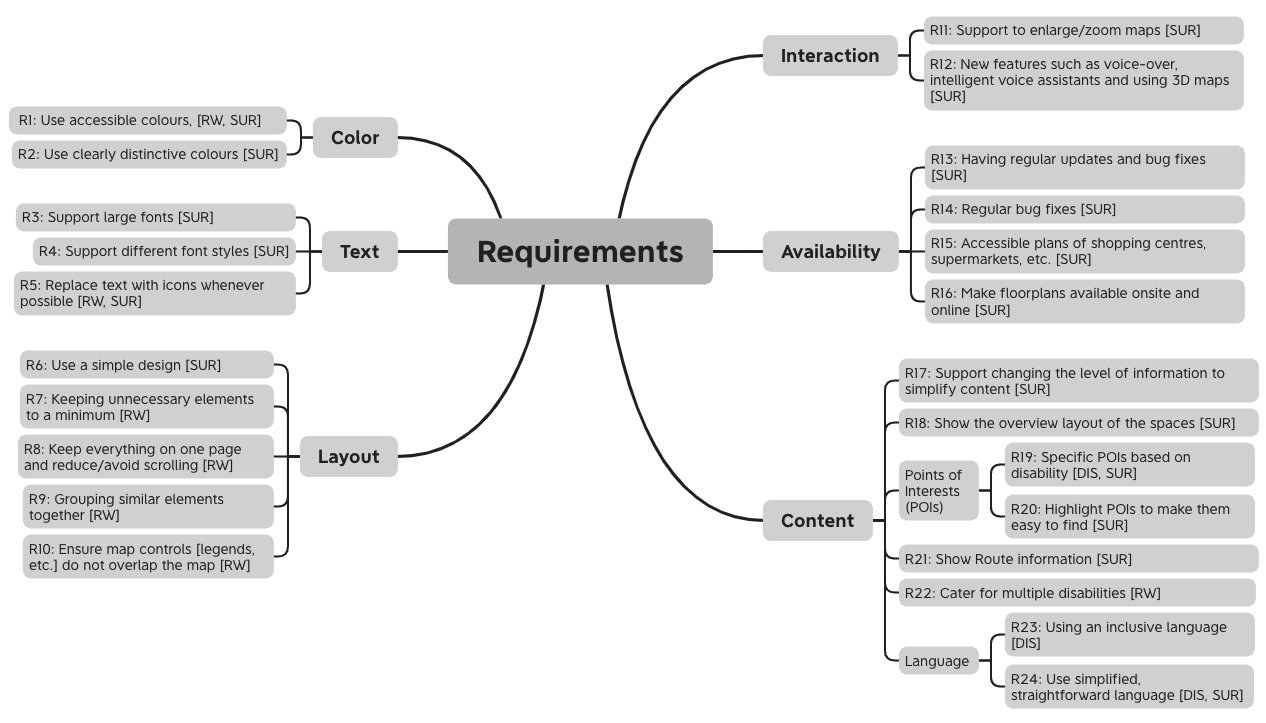}
\caption{Requirements from Related work [RW], Discussion [DIS] and Survey [SUR]}
\label{fig_reqs}
\end{figure}

The requirements we identified from our three-step process are shown in Figure \ref{fig_reqs}. These were categorised into four categories: Color, Text, Layout, Interaction, Availability and Content. As shown in the figure, each of these requirements are numbered as R1...Rn. We will use these numbers as \textbf{[Rx]} to refer to specific requirements when we explain our design decisions. 

\section{Prototype development}
\begin{figure}[h]
    \includegraphics[width=\linewidth]{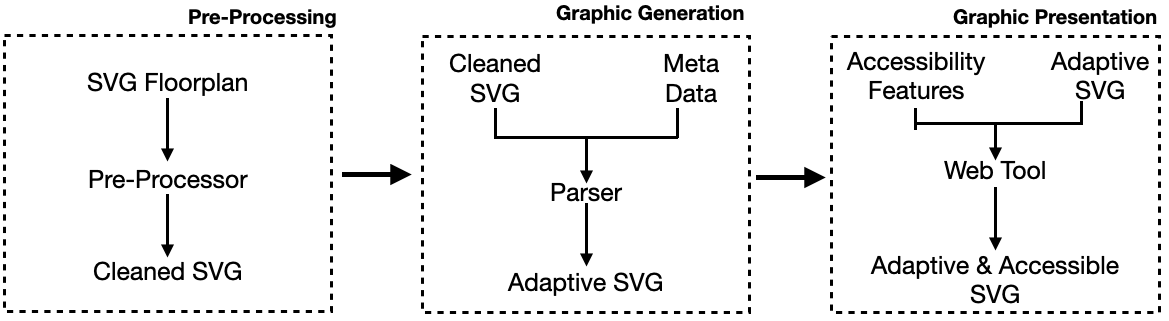}
    \caption{Adaptive Graphic Generation Framework}
    \label{fig_framework}
\end{figure}

We created a workflow, as shown in Figure \ref{fig_framework}, to generate accessible and adaptive graphics and implemented it with an SVG map of a large retail centre from a university \textbf{[R15]}. 

\subsection{Pre-processing}
SVG files have an XML-based back-end, which allows direct manipulation of the graphic without causing any distortions on the visual aspect. However, SVG can be created by different tools, e.g. Inkscape, and CAD. Based on the used tool and the creator's design decisions, the XML can be simple or quite complicated \cite{madugallagenerating}. Therefore, this step focuses on cleaning the SVG to remove unnecessary data layers to generate a cleaned SVG as shown in Figure \ref{fig_framework}.

\subsection{Graphic Generation Stage}
This step accepts the cleaned SVG and any metadata associated with that file as input. For example, in generating an accessible university map, we received SVGs of building plans for the retail centre and metadata about rooms in them from the Monash Building and Planning department. These metadata included room names, categories, and departments/faculties to which the rooms belong. In this scenario, the metadata was not incorporated into the original SVG as it would make the SVG’s XML too complicated. We coded our parser in the Python programming language, and it performed two main tasks. Task 1 was combining the SVG with metadata (currently in CSV format) and generating a more detailed SVG. Task 2 was adding pre-defined adaptive features to the new SVG. The output of the Parser was an adaptive SVG. The pre-defined adaptive features that we added to the SVGs are listed below. 
    
\begin{enumerate}
    \item HTML elements that support accessibility: HTML usually acts as the container for SVG, with other web technologies such as JavaScript and CSS providing additional support for SVG presentation. Most HTML and SVG elements support accessibility features such as tooltips and screen readers. However, this is not true for all HTML and SVG elements \cite{R15}. Therefore, we conducted a preliminary evaluation across web browsers and screen readers to find the elements that work best and included these in our accessible SVG structure, e.g. \textit{<aria-describedby>} chosen over \textit{<desc>} tag, as it was compatible with all screen readers \textbf{[R12]}.
    \item HTML Focus Feature: HTML focus refers to which item on the screen (a button, menu, text) currently receives input from the keyboard. The screen readers and tooltips will present the information in ‘HTML Focus’ when interacting with HTML pages. We used this feature to help screen readers access Accessibility layer-based information \textbf{[R12- support for screenreaders]}.
    \item Colours: We used IBM Design’s colour blind safe palette \cite{R21} as the basis for our solution and generated a half-fill colour mechanism to increase the number of colour options as shown in Figure \ref{fig_fill_pattern} \textbf{[R1,R2]} We also supported allowing users to change these colours at run time to suit their personal preferences.
    \item Patterns: According to WCAG 2.0, Guideline 111, when colour differences are used to convey information, patterns should optionally be available to depict the same information \cite{R26}. Therefore, in our adaptive graphic structure, we added support for patterns as shown in Figure \ref{fig_fill_pattern}. 
    \begin{figure}[h]
        \centerline{\includegraphics[width=0.6\columnwidth]{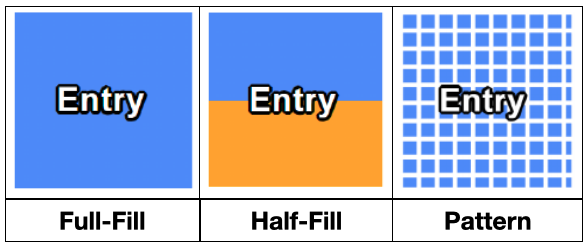}}
        \caption{Full-fill and half-fill colours and Pattern fill}
        \label{fig_fill_pattern}
    \end{figure}
    \item SVG Layers and Bit Fields \label{section_layers}: We use SVG layers to provide adaptive information based on disabilities. Our Accessibility Layers represent each disability, and switching them on or off helps to present building information based on disabilities, e.g. for the mobility impaired, the stairs are hidden, and elevators are highlighted. To support users with multiple disabilities, our layered architecture allows these layers to be overlapped intelligently. We used the Bit field data structure, which is popular in the C Programming language, to switch on and off these layers \textbf{[R17,19,20,22]}. 
    
    
    
\end{enumerate}

\begin{figure}[h]
\centerline{\includegraphics[width=\textwidth]{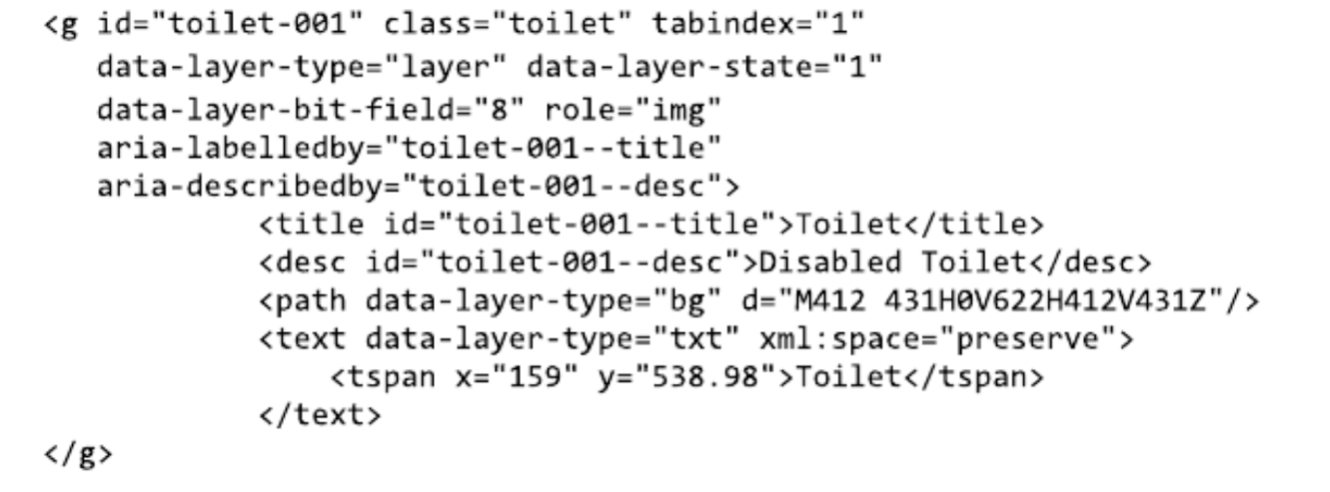}}
\caption{XML structure for an Adaptive SVG for a Toilet in a Floorplan}
\label{fig_adaptiveXML}
\end{figure}

After integrating these features into SVG, we could generate an adaptive SVG. In Figure \ref{fig_adaptiveXML}, we have given an extraction of this adaptive SVG, and it presents the use of ``data-layer-bit-field" for disability-based layers, 'data-layer-state' to show activated/de-activated layers, ``tab-index" for focus and <title>, <desc> for element details. 

\subsection{Graphic Presentation}
This step involved developing the web tool, which accepted the adaptive SVG created in the Graphic Generation stage. 

\subsubsection{Design and Structure: }
\begin{figure}[h]
 \subfloat[Main UI \label{fig_mainUI}]{%
    \includegraphics[width=\textwidth]{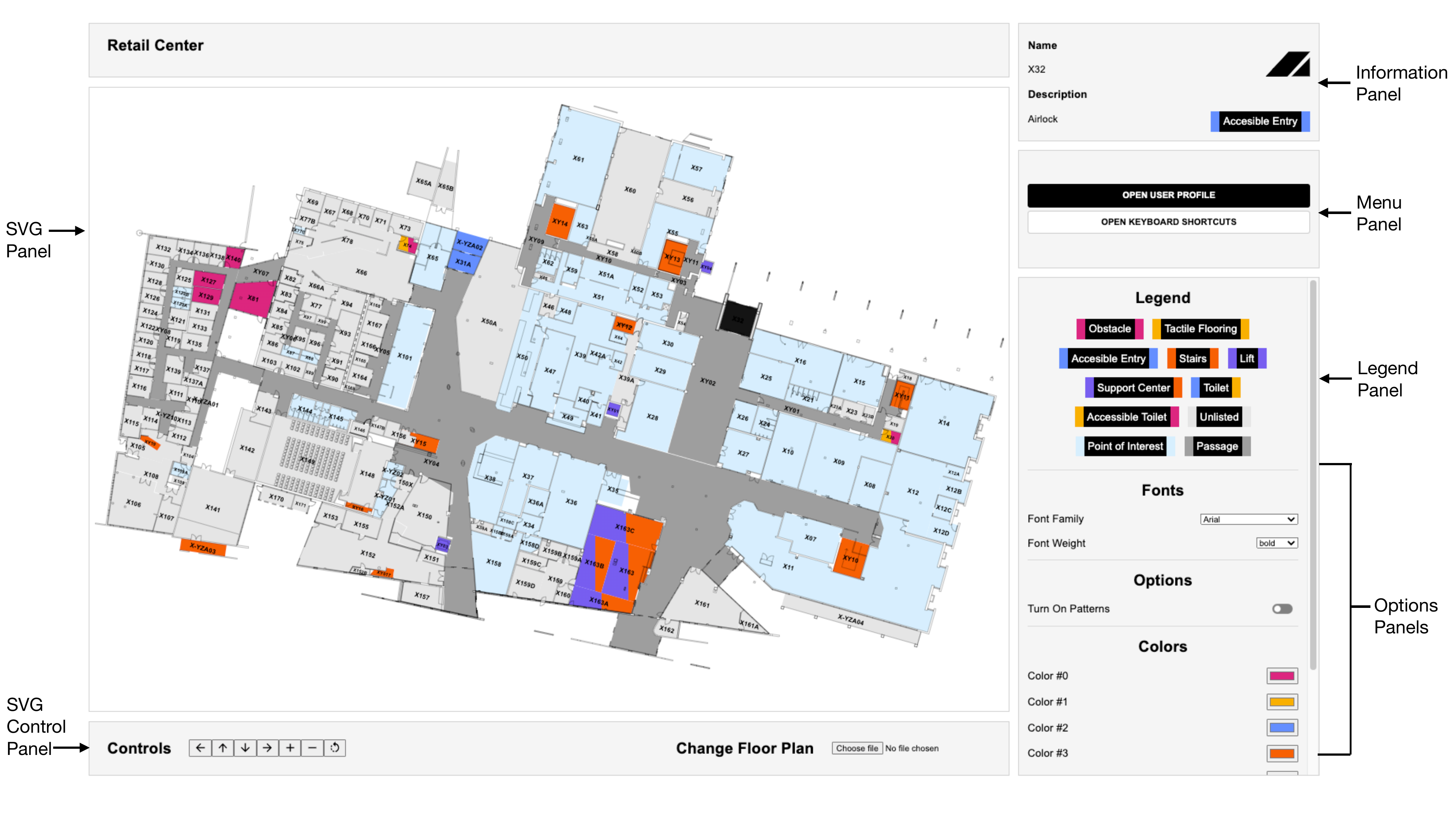}}\\
  \subfloat[User Profile Menu \label{fig_UserProfile}]{%
      \includegraphics[width=0.6\textwidth]{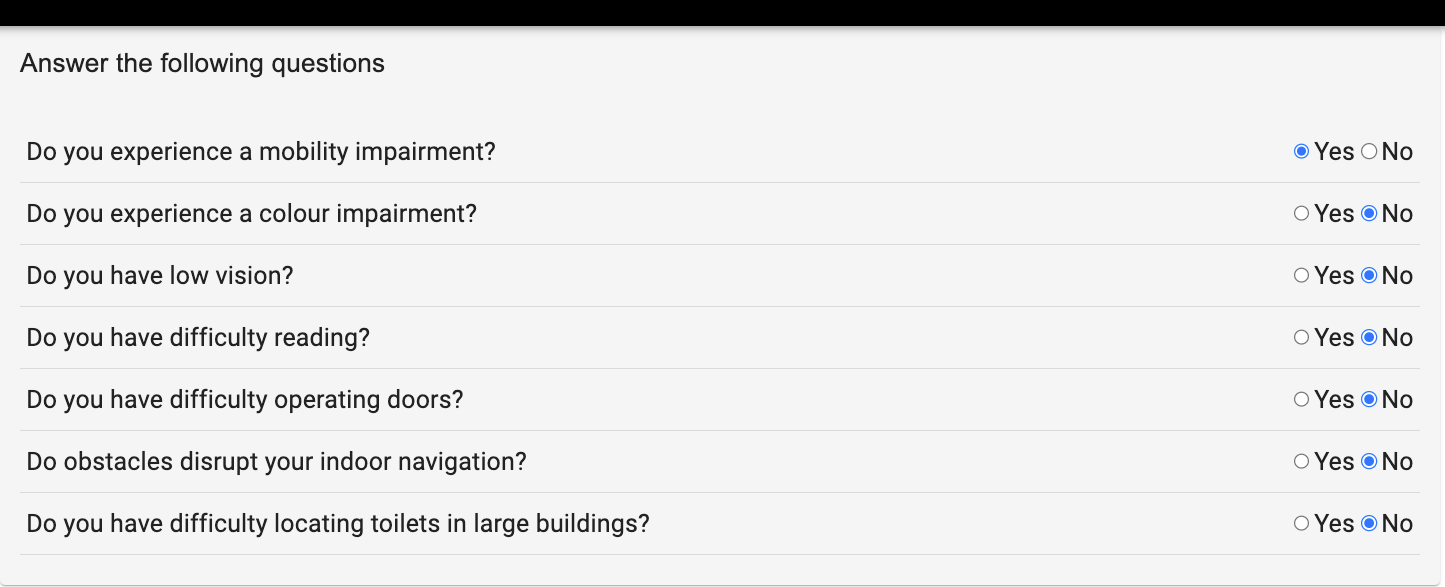}}
    \hfill
  \subfloat[Keyboard Shortcuts \label{fig_Keyboard}]{%
        \includegraphics[width=0.38\textwidth,trim={2cm 3cm 2cm 4cm},clip]{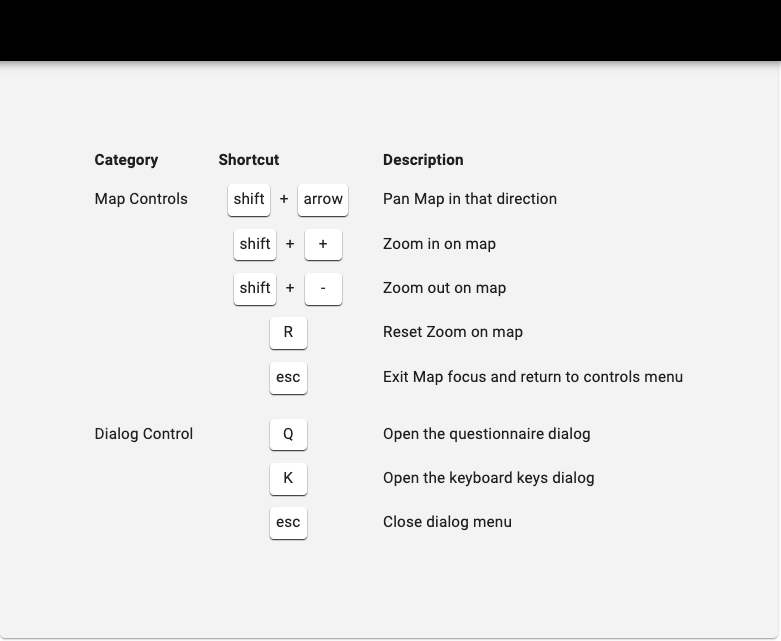}}
  \caption{The web tool\\ Available at :\url{https://accessible-svg.github.io/floorplan-viewer/}}
  \label{fig_WebToolUI} 
\end{figure}

To address requirements, we designed our tool with three main pages, as shown in Figure \ref{fig_WebToolUI}. Page 1 was the main screen with the map (Figure \ref{fig_mainUI}), page 2 was the ``User profile" page (Figure \ref{fig_UserProfile}), and page 3 was for the keyboard shortcuts (Figure \ref{fig_Keyboard}). We designed the UI of page 1 in a grid layout, with the different panels grouping similar key features \textbf{[R9]}, and we took steps to use a simple and inclusive language [\textbf{R23, 24}]. A user profile can be created by clicking 'Open User Profile' and moving to page 2. The shortcuts can be found by clicking the ``Open Keyboard Shortcuts' and opening page 3. 

We implemented this tool using the React framework \textbf{[R16]}. It was based on a use case of the Monash University building map provided by the Monash University Building and Planning Department. In the tool, to explore the adaptive features, a user must create a profile with disability information as shown in Figure \ref{fig_UserProfile}. This will enable/disable certain SVG layers, and it will lead to a version of the floorplan that is adapted to fit the specified disability \textbf{[R17, 22]}). The UI of this web tool contained five simple components, and these are listed below and are shown in Figure \ref{fig_mainUI} \textbf{[R6, 7, 8, 10]}.


\begin{enumerate}
    \item SVG Panel: Assigned the largest block on the UI and shows the graphical view of the SVG. Users can interact with this graphic by clicking on elements and using the control panel below \textbf{[R18]}.
    \item SVG control Panel: Contains features to control the floor plan, like the panning and zooming buttons or the ability to load in a new SVG \textbf{[R11]}.
    \item Information Panel: Whenever a graphical element is selected with either the keyboard or the mouse, the element’s title, description, and any associated icons describing the element are displayed \textbf{[R5]}.
    \item Menu Panel: Contains the buttons to open the dialogue menus, define user disabilities, and explore keyboard shortcuts. We added these under separate screens to follow guideline 3 of \cite{R16} and reduce scrolling. 
    \item Legend Panel: The legend contains different element types and their assigned colour details. Any changes made to the colour options below are reflected in the legend as well.
    \item Options Panel: Groups different options to tweak the colour, font, or apply patterns to the floor plan \textbf{[R1,2,3,4]}.
\end{enumerate}

\subsubsection{Using the Web Tool: }
To ensure all UI components were accessible via keyboard and screen readers, we used an aria-specific react library for elements like drop-down menus, radio buttons, tables, and switch buttons. To make the legend panel accessible for screen readers, we added tab indexes. Similarly, the control buttons were embedded with additional aria-label tags to describe their function, as they are icon-based buttons. The information panel displayed the information related to any element selected via keyboard or mouse. In using this tool, a user will first answer the user profile questionnaire, which will lead to activating necessary SVG layers and adapting the map accordingly. A user with multiple disabilities can answer ``Yes" to more than one question, and this will lead to activating multiple layers as needed.

\section{Evaluation}
After the framework was completed, we evaluated the adaptive SVG and the Web tool, as these were the components of our framework that involved direct interactions with disabled users. For this, we conducted semi-structured interviews with seven disabled users to get feedback from potential users. 

\subsection{Design and Recruitment }
In this evaluation, we requested our participants to explore our adaptive SVG-based floorplans via our web tool and evaluate them. This involved first playing an introductory video of the web tool and answering any clarification questions participants may have to give them an idea of how to interact with it. All participants found the video helpful in understanding the tool's functionality. Then, we provided a task list, as shown in Table \ref{tab_tasks}, and asked the participants to perform these tasks by interacting with our prototype tool. We then interviewed participants and each interview took approximately 45 minutes.

\begin{table}[h]
\caption{Tasks to interact with web tool}
\begin{tabularx}{\columnwidth}{|X|} 
 \hline
 \textbf{Task 1}: Open ``User profile", and select options that suit the participant\\
 \hline
 \textbf{Task 2}: Open the user profile and answer "yes" to “Do you experience a mobility impairment?”. You'll see that accessible entries are represented with blue, as seen on the legend and the map. Change the colour of the accessible entries to black.  \\ 
 \hline
 \textbf{Task 3}: Change the font of the text labels to OpenDyslexic using tool functions\\
 \hline
 \textbf{Task 4}: Can you download the SVG at \href{https://fit4003-group19.s3.ap-southeast-2.amazonaws.com/pattern--demo.svg}{this link}. Now, can you switch the floor plan to the new SVG?\\
 \hline
 \textbf{Task 5}: Return to the original floorplan by clicking refresh. Turn on the ``pattern mode" in the web tool and identify three elevators \\
 \hline
 \textbf{Task 6}: Starting at the Support Center (labelled by G163), can you identify a route to the nearest Accessible Toilet?\\
 \hline

\end{tabularx}
\label{tab_tasks}
\end{table}

We advertised our study via social media platforms, the university disability support centre, and personal contacts. We received seven responses and conducted Zoom-based online interviews with them. 

\subsection{Data Analysis}
We audio-recorded each interview and transcribed these later using Zoom's transcription feature. We used a mixed-method approach to analyse the data. We had four demographic questions, six tasks, feedback on these tasks and questions on general feedback for UI. We conducted descriptive statistical analysis on demographic questions and task completion. The feedback on tasks and general UI were analysed using thematic analysis.

\subsubsection{Demographics}
The results of the demographics analysis are shown in Table \ref{tab_intr_demographics}. We had seven participants and at least one representative from our target participant groups. 

\begin{table}[h]
\caption{Demographics of Interview Participants}
\begin{tabular}{l r | l r | l r | l r} 
 \toprule
 \textbf{Gender} &   & \textbf{Age} &  & \textbf{Disability} &  & \textbf{Duration } & \\ 
 \midrule
 Male & 2 & \textless 30 & 4 & Low vision & 3 & \textless 5 yrs. & 1\\
 Female & 5 & 30+ & 3 & Wheelchair  & 1  & 5-10 yrs. & 1 \\
  & & & & Colour Impaired & 1 & 11-20 yrs. & 2 \\
  & & & & Motor Impaired & 2 &  21+ yrs. & 3\\
 \bottomrule
\end{tabular}
\label{tab_intr_demographics}
\end{table}

\subsubsection{Task Completion \newline}
The participants were given six tasks, as shown in Table \ref{tab_tasks}, and most of our participants completed these six tasks successfully. The ease of completion of each task is shown in Figure \ref{Evaluation} below, with different colours indicating participant disabilities. 

\begin{figure}[h]
\includegraphics[width=\textwidth, trim={0.2cm 0.2cm 0.2cm 1cm},clip]{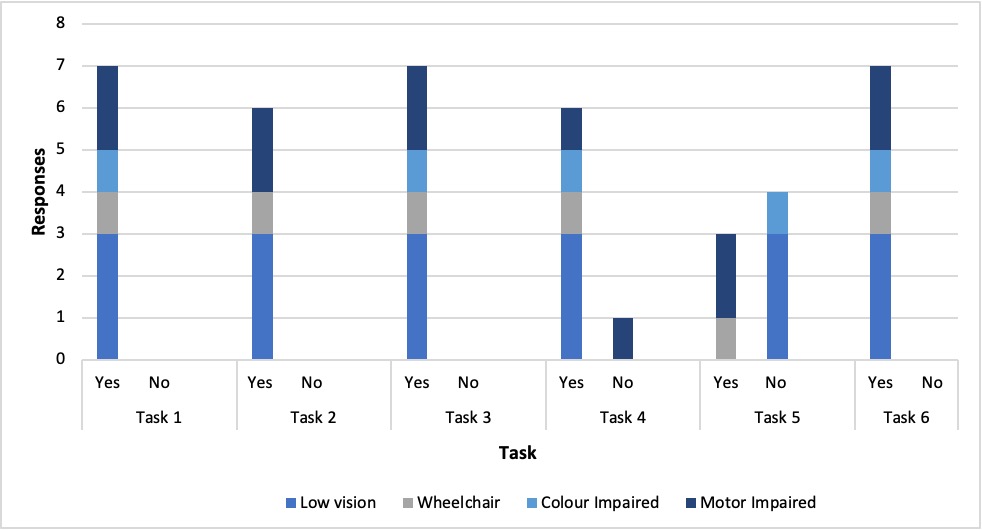}
\caption{Ease of task completion for interview participants} 
\label{Evaluation}
\end{figure}

In \textbf{task 1}, our participants mentioned that the user profile function was easy to find, and the impairment-related selections were easily accessible. However, a participant with low vision suggested making the user profile button more significant and evident. 
For \textbf{task 2}, six participants completed the task quickly, and one participant could not complete the task. Those who completed the task easily mentioned that they had no problem changing colours for different legends. However, the participant who failed to complete the task was colour-impaired; he stated that he ``could not identify similar colours (blue/purple) when they are used to represent different elements". 
For \textbf{task 3}, all participants completed the task easily; they stated that the font button was easy to find, and all fonts were recognisable and reader-friendly. 
For \textbf{task 4}, six participants completed the task easily, and one participant found it moderately difficult to complete the task. Those who completed it easily stated that the procedures of the whole task were straightforward, while the motor-impaired participant who found this task difficult felt that there were too many steps and that the process was tiring. 
For \textbf{task 5}, three participants completed the task easily, and four participants found it moderately difficult to complete the task. However, those who completed it easily stated that the locations were easy and obvious to find. Alternatively, the reasons for the moderate difficulty were: 1) the colours representing certain locations were overlapping, 2) the ``pattern mode" of the web tool was difficult to use, and 3) Certain locations were too small on the web tool, making them confusing to find. The participants who struggled with this task were three vision-impaired participants and one colour-impaired participant, suggesting that the patterns were hard to use when you had a vision issue. 
For \textbf{task 6}, all participants completed the task easily; they identified multiple routes on the floor plan and chose the shortest route or the first route seen. Apart from the one participant who could not complete task 2, all participants completed the other tasks with ease or moderate difficulty.

\subsubsection{Feedback on UI and SVG \newline}
\textbf{Feedback: }We asked our participants about their thoughts on the web tool UI, and all participants said that the UI was ``easy to use", ``easy to follow", and ``tidy and skilful". We also asked them if the web tool provided help to them, and all seven of our participants said yes. The reasons for yes were that 1) the web tool helped them find different locations, 2) the legends with different colours were helpful, and 3) the tool was convenient and easy to use. All our participants with mobility issues liked the tool’s features, highlighting accessible toilets and entries using different colours. They mentioned that they would always use the floor plans to look for lifts and accessible toilets when they go out, and this web tool made it easier to find these: ``The legends were very helpful with places like lifts, toilets, and stairs, the colours made it easier to distinguish different areas". We can conclude that our participants were generally satisfied with the web tool.

\textbf{Improvements: }Regarding suggestions for improvement, some of our colour-blind participants suggested using a brighter colour palette like blue or orange. They recommended refraining from using similar colours like blue and purple as it makes separating elements with similar colours difficult. Six of our participants felt the pattern mode of the web tool was difficult to use and ``hard to play with". They thought it made the text labels hard to read at times. One participant suggested making the User Profile button more apparent, and one suggested adding hover text for icon buttons in the SVG control panel. To help with navigation, they suggested displaying the shortest routes with bright colours and highlighting accessible routes with ramps and accessible parking areas.

\section{Threats and Limitations}
In our requirement gathering survey, we translated the survey and the responses to Mandarin and back when gathering responses from the eye hospital and the disability school in China. We ensured that the ethics committee was aware of this, and several team members went through responses after receiving the translated responses. But we acknowledge that even with these measures, minor mistakes may happen in translations. 

Another threat is in the evaluation of the tool. We had only seven participants in this evaluation, and while they have significant distribution between disabilities, a more detailed evaluation would be beneficial. It may also be beneficial to obtain feedback from the experts in this stage, similar to the requirement-gathering stage. 

Our work presented a detailed evaluation from the point of view of disabled end users. However, it would also be beneficial to obtain feedback from front-end developers who would be the ones using our adaptive graphic generation framework. Therefore, future studies can focus on getting feedback from font end developers on the structure of the adaptive SVGs and the adaptive graphic generation framework. 

This study focused on four types of most common disabled groups. While our technical framework will be able to address more types of disabilities, a new requirement-gathering exercise will need to be conducted for these other disabilities. More work is also needed to solve the situation where these disabilities conflict with each other, e.g., one obscures the other, or the overall amount of information becomes excessive. It can also explore integrating the shortest path-finding function into the tool, as this is critical in navigating complex public spaces. 

We chose maps for our case study as they had a significant real-world impact on disabled communities. But there are other types of information graphics, such as charts and diagrams, that are used more in software applications. Future work can, therefore, explore using our framework and guidelines to generate accessible and adaptive versions of other types of information graphics. 

Lastly, while our graphic generation framework provides an excellent one-stop solution to support multiple disabilities, it is still at the proof of concept stage. Therefore more work is needed to explore the possibility of standardising some of these adaptivity aspects so that they are not vendor or tool-dependent.




\section{Recommendations}

\textbf{Implications for front-end developers:} 
Front-end developers can utilise our findings to aid them in engineering graphics and other online content accessible to people with disabilities. While this paper focuses on developing adaptive public space plans, the approach can be used to make different types of information graphics such as charts (bar, pie), diagrams (Venn, flow, mind maps), other maps (political, thematic, public space) that are used in different domains such as education, health, finance, legal, defence, entertainment more accessible. Some guidelines to be followed in engineering graphics for special groups are shown in Table \ref{tab_guidelines}.

\begin{table}[h]
\caption{Guidelines in Engineering Graphics for Disabled}
\begin{tabularx}{\columnwidth}{| c | X |}
 \hline & \\[-2ex]
   \multirow{4}*{\Huge \faBlind} & Provide sufficient contrast using colours and patterns. In using patterns, ensure they are simple and do not overlap with text\\
    \cline{2-2} & \\[-2ex]
     \multirow{0}*{Low} & Always provide the ability to manually zoom in/out without causing any distortion to graphics\\
     \cline{2-2} & \\[-2ex]
     \multirow{-1}*{Vision}& Provide support for screen readers and keyboard accessibility\\
    \hline & \\[-2ex]
    \multirow{3}*{\Huge \faLowVision}  & Use colour impaired-friendly colours \\
    \cline{2-2} & \\[-2ex]
    \multirow{4}*{Color} & Ensure there is high contrast, or use monochrome colours (using multiple shades of a single colour)\\
     \cline{2-2} & \\[-2ex]
    \multirow{-0.5}*{Impaired} & Do not place reliance only on colour to convey information; instead, combine it with text and symbols\\
    \hline & \\[-2ex]

    \multirow{4}*{\Huge \faGroup}  & Use icons instead of labels as much as possible\\
    \cline{2-2} & \\[-2ex]
    \multirow{5}*{Dyslexia} & Keep the graphic simple and, if required, present it in multiple layers\\
    \cline{2-2} & \\[-2ex]
     & Use dyslexia-friendly fonts when possible\\
    \cline{2-2} & \\[-2ex]
     & Use a simple language\\
    \hline & \\[-2ex]

    \multirow{3}*{\Huge \faWheelchair}  & Graphics such as maps require, \\
    \multirow{4}*{Mobility} & - Locations of ramps, lifts, accessible toilets\\
    \multirow{4.5}*{Impaired} & - Paths with wider corridors\\
     & - Sufficient turning cycles\\
     & - Less crowded areas\\
    \hline
\end{tabularx}
\label{tab_guidelines}
\end{table}



\textbf{Implications for public spaces owners:} Our requirement gathering survey has detailed information on types of public spaces used by disabled users, frequency of use,  methods they use to access information about these spaces, issues and suggestions. Public space owners can use our data to understand how disabled groups consume their spaces and take steps to make these more accessible by trying to solve the issues they have reported and by integrating some of their suggestions. Such initiatives will create satisfied consumers who will visit these spaces more and may recommend them to peers, e.g. a wheelchair user who found it easy to locate lifts and ramps at retail would recommend it to other wheelchair users. This would give a competitive advantage to these public space owners over others. 

\textbf{Implications for end users:} The described case study will help make public spaces more accessible for disabled users, helping them to pre-plan their visits to these spaces and navigate independently once inside. Since our approach can be used to make other types of graphics accessible as well, it will contribute to creating more graphics accessible for disabled users. 


\section{Summary}
In this study, we propose an approach used to generate adaptive and accessible graphics for people with multiple disabilities, specifically for low vision, colour blindness, dyslexia and mobility impairment. Our research consisted of a requirements elicitation stage involving informal interviews with accessibility experts and surveys with disabled users. After analysing these requirements, we developed a framework to generate adaptive SVG-based graphics and applied this using a case study of complex indoor environment floorplans. Finally, we evaluated our approach by using a preliminary evaluation with seven disabled users. The approach shows potential for creating more accessible graphics for various disabled users using adaptive SVG-based graphics.

\section{Declarations} 
\textbf{Conflict of Interests:} The authors declared that they have no conflict of interest.

\textbf{Data Availability Statement:} The data sets generated during and/or analysed during the current study are available from the corresponding author at reasonable request.


%
%

\bibliographystyle{spmpsci}      
\bibliography{References.bib}

\appendix

\end{document}